\documentclass[reprint,showpacs,preprintnumbers,amsmath,amssymb,prl]{revtex4-1}
\usepackage{amsmath}
\usepackage{amsfonts}
\usepackage{amsthm}
\usepackage{dsfont}
\usepackage{graphics}
\usepackage{graphicx,bm}
\usepackage[caption=false]{subfig}
\usepackage{amssymb}
\usepackage{mathrsfs}
\usepackage{braket}
\usepackage{float}
\usepackage{color}
\usepackage{url}
\usepackage[abs]{overpic}
\usepackage[usenames,dvipsnames]{xcolor}
\usepackage{commath}
\usepackage{subfig}
\usepackage{multirow}
\usepackage{verbatim}
\usepackage{array}
\usepackage[hidelinks]{hyperref}

\begin{document}

\title[
Minimal quantum walk simulation of Dirac fermions in curved space-times]{Minimal quantum walk simulation of Dirac fermions in curved space-times} 

\author{Fabrice Debbasch$^{1}$}
\email{fabrice.debbasch@gmail.com}

\affiliation{
$^{1}$Sorbonne Universit\'e, Observatoire de Paris, Universit\'e PSL, CNRS, LERMA, F-75005, {\sl Paris}, France}

\date{\today}
\begin{abstract}
The problem of simulating through quantum walks Dirac fermions in arbitrary curved space-times and coordinates is revisited, taking $(1 + 1)$D space-times as an example. A new shift or translation operator on the grid is introduced, to take into account arbitrary geometries. The new, generalised quantum walks built with this operator can simulate Dirac fermions in arbitrary curved space-times and coordinates, and their wave functions have exactly the same number of components as standard Dirac spinors, and not twice that number, as previously believed. In particular, in $(1 + 1)$D space-times, only one qubit is needed at each lattice point, which makes it easier to perform quantum simulations of the Dirac dynamics on current NISQs quantum devices.
Numerical simulations of the Dirac dynamics in the post Newtonian, so-called Gravitoelectromagnetism regime are presented as an illustration.


\end{abstract}
\keywords{gggggg}
\maketitle

%
%
%

\section{I. Introduction}

Dirac fermions are spin $1/2$ particles.  They are represented in quantum theory by spinors which obey the Dirac equation. This equation was introduced by Dirac in 1928 for free fermions propagating in flat space-time and later extended to describe the coupling of Dirac fermions with arbitrary interaction fields, be they Yang-Mills gauge fields and or gravitational ones. 
The quantum simulation of Dirac fermions
 has been an active topic for more than ten years, with applications ranging from fundamental physics to astrophysics and condensed matter. Quantum simulation can be achieved through Quantum Walks (QWs), which are automata defined on graphs and lattices. They were first considered by Feynman in studying possible discretisations for the Dirac path integral \cite{feynman2010quantum,schweber1986feynman}. They were later introduced in a systematic way by Aharonov et al. \cite{aharonov1993quantum} and Myers \cite{meyer1996quantum}.
QWs are useful in quantum information and algorithmic development \cite{ambainis2007quantum,magniez2011search,ManouchehriWang2014} because they are a universal computational primitive.
They 
have also been realised experimentally in a number of ways \cite {ManouchehriWang2014} which include cold atoms \cite{Ketal09a}, photonic systems \cite{Petal10a,Setal10a} and trapped ions \cite{HAetal20a}. 

QWs can be used to simulate, not only free Dirac fermions, but also Dirac fermions interacting with arbitrary Yang-Mills gauge fields \cite{arnault2016quantum,marquez2018electromagnetic} and arbitrary relativistic gravitational fields \cite{di2013quantum, di2014quantum, AFF16a,AF17a,arnault2017quantum,ADMMMP19a}. Quite remarkably, the QW-based models incorporating Yang-Mills fields display exact discrete gauge invariances which can be used to build exact discrete Yang-Mills field strengths \cite{arnault2016quantum2}, and one can even construct a discrete counterpart to the Riemann curvature tensor \cite{D19b}. Also, discussions of the Lorentz invariance of QWs can be found in \cite{AFF14a,BDAP17a,D19a,ABDAP19a} while the symplectic discrete geometry behind some QWs has been presented in \cite{D19a}. And even Non Commutative Geometry can be simulated with QWs \cite{Debbasch21a}.
 
 If the construction of QWs simulating Dirac fermions in flat space-time and orthonormal, so called Minkovski coordinates is relatively straightforward, even if one includes Yang-Mills gauge fields, the construction of so-called Gravitational Quantum Walks (GQWs) {\sl i.e.} QWs simulating Dirac fermions in curved space-times (or in flat space-time, but  with arbitrary coordinates)is much more complex. 
 
 Three kinds of GQWs have been proposed, each with its own apparent advantages and drawbacks. 
 On the positive side, all proposed GQWs can simulate Dirac fermions in arbitrary gravitational fields. The problem lies with the choice of grid and coordinates. In essence, the only known GQWs which can simulate fermions on a simple square grid in arbitrary coordinate systems use wave-functions which have twice as many components as the corresponding Dirac fermions in continuous space-times. The other GQWs do not double the number of components, {\sl i.e.} they use wave-functions with the same number of components as the continuous space-time Dirac fermions, but they are restricted in the choice of grid and/or coordinate systems and/or space-time dimension. 
 
 This current, quite unsatisfying situation with GQWs may seem anecdotical, but it is actually an instance of a much larger and definitely not anecdotical issue {\sl i.e.} discretising continuous physical problems involving non trivial geometry, be it for computational or purely theoretical purposes. For example, continuous physics uses differential geometry to define and deal with geometrical notions 
like curvature and parallel transport. But differential geometry is by definition local in space and time and does not provide any clue on how to define transport between points on a discrete grid. 
 
The aim of this article is to propose new, generalised GQWs which eschew all the shortcomings of the previously proposed GQWs by shedding a new light on how to achieve unitary, point-dependent transport on a discrete grid, thus making it possible to take into account arbitrary local geometries in a discrete setting. As previous GQWs, the new GQWs
can simulate Dirac fermions propagating in space-times of arbitrary dimensions with arbitrary gravitational fields. But they are defined on simple square grids which discretise arbitrary space-time coordinates and their wave functions have the minimal number of components equal to the number of components of Dirac spinors. There are two basic steps in the construction of these generalised walks. The first one is to put the Dirac equation into a quantum computing friendly form. The second one is to generalise the standard shift operators used in defining standard quantum walks, to accommodate arbitrary local choices of $n$-bein in possibly curved space-times of arbitrary dimensions. 

For simplicity sakes, this article focuses on the simplest case possible {sl i.e.} space-times of dimension $2$. Section II presents the so-called curved space-time Dirac equation and puts it into 
a form suitable for quantum simulation. Section III presents the construction of the generalised quantum walks and Section IV offers a detailed analytical and numerical study of generalised walks propagating in post-Newtonian, so-called gravitoelectromagnetic \cite{Wald84a,Mashhoon08a} space-times. All results are finally discussed in Section V.

The overall conclusion is that the minimal quantum simulation of the Dirac equation generically involves quantum walks with wave-functions which have as many components as the usual Dirac spinors, not more. In particular, in dimension $(1 + 1)$, the propagation of Dirac fermions in arbitrary coordinates can always be simulated by a 
single qubit at each lattice point, and not $2$ as in GQWs of the second kind (GQWs of the third kind simply do not exist in $(1 + 1)$D). Naturally, keeping the qubit number minimal is of paramount importance for NISQ-based simulations, which are the only quantum simulations accessible today.

\section{II. Dirac dynamics in $(1 + 1)$D spacetime}

\subsection{II.1 Dirac equation}

The Dirac equation in space-times of arbitrary dimension reads
\begin{equation}
e^\mu_a \gamma^a \nabla_\mu \Psi + i m \Psi = 0.
\end{equation}
Here, $\Psi$ is a spinor field and $m$ is its mass. The quantities $e^\mu_a$ are the $n$-bein coefficients, which obey
\begin{equation}
g_{\mu \nu} e^\mu_a e^\nu_b = \eta_{ab}
\end{equation}
where $\eta_{ab}$ are the components of the flat Minkovski metric in orthonormal 
coordinates and $g_{\mu \nu}$ are the 
coordinate components of the arbitrary space-time metric. 
By convention, the signature of all metrics will be $(+, -, ... , -)$. The $\gamma$ operators obey
the standard, flat space-time Clifford algebra
\begin{equation}
\left \{ \gamma^a, \gamma^b \right\} = \eta^{ab},
\end{equation}
where $\eta^{ab}$ is the inverse of $\eta_{ab}$. The $n$-bein coefficients can also be caracterized by
\begin{equation}
\eta^{ab} e^\mu_a e^\nu_b = g^{\mu \nu}.
\label{eq:defnbein}
\end{equation}
Let us stress that the Clifford algebra does not define a unique set of $\gamma$ operators. Similarly, since both the metric $g$ and the $\eta$'s are symmetric, 
condition (\ref{eq:defnbein}) gives less independent equations than there are $n$-bein corefficients and, thus, does not define a unique $n$-bein.
For example, in $(1 + 1)$D space-time, (\ref{eq:defnbein}) transcribes into the 3 independent equations
\begin{eqnarray}
(e^0_0)^2 - (e^0_1)^2 & = & g^{00} \nonumber \\
e^0_0 e^1_0 - e^0_1 e^1_1 & = & g^{01} \nonumber \\
(e^1_0)^2 - (e^1_1)^2 & = & g^{11}.
\label{eq:2beinrel}
\end{eqnarray}
and these are not enough to fix the 4 $2$-bein coefficients unambiguously in terms of the metric components. There is therefore a substantial freedom in the choice of $2$-bein
and, more generally, of the $n$-bein. This freedom will be used in the next Section to put the Dirac equation into a form suited for quantum simulation by minimal GQWs. Without going details at this stage, let us only remark that (\ref{eq:2beinrel})  being quadratic makes it possible to always impose the conditions $e^0_0 \ge 0$ and $e_1^1 \ge 0$ on the $2$-bein. This is so because, since (\ref{eq:2beinrel}) is quadratic in the $2$-bein coefficients, if $(e_0, e_1)$ is a $2$-bein, so are $(-e_0, e_1)$, 
$(e_0, -e_1)$ and $(-e_0, -e_1)$. 
Now, since the coordinate $x^0$ is time-like, $g^{00}$ is necessarily strictly positive and the first relation in (\ref{eq:2beinrel}) implies $\mid e^0_0 \mid > \mid e^0_1 \mid$. It follows that $e_0^0$ is strictly positive, and also that $e^0_0$ is strictly superior to both $e^0_1$ and $- e^0_1$. Similarly, $x^1$ being space-like implies that
$g^{11}$ is strictly negative, so $\mid e^1_1 \mid > \mid e^1_0 \mid$ and the coefficient $e_1^1$ is therefore also strictly positive. From all this follows that the quantities 
$\sigma_0$ and $\delta_0$ introduced in the next section never vanish.

We finally note that requiring $\eta_{ab}$ to be identical to the components of the Minkovski metric in orthonormal coordinates precludes using null $n$-beins. This 
does not restrict the gravitational field or the space-time coordinates, but rather the tools used to write the Dirac equation. 

The operator $\nabla_\mu$ is the covariant derivative of the spinor $\Psi$. Given a basis
$(b_\sigma)$ is spinor space, the spinor $\Psi$ is represented by its components $\Psi^\sigma$
and components of the spinor $\nabla_\mu \Psi$ are:
\begin{equation}
(\nabla_\mu \Psi)^\sigma = \partial_\mu \Psi^\sigma + \frac{i}{8}\, \omega_{\mu a b} \left( \left[ \gamma^a, \gamma^b \right] \Psi \right)^\sigma
\end{equation}
where the $\omega$'s are the so-called spin rotation coefficients.
Since the connection is compatible with the metric, the $\omega$'s 
depend on the metric coefficients and on their partial derivatives or, alternately, on the $n$-bein coefficients and their partial derivatives.
In $(1 + 1)$D space-times, it is convenient to mix both options and the Dirac equation obeyed by the two-component spinor $\Psi$ can be written as:
\begin{equation}
e^\mu_a \gamma^a \partial_\mu \Psi + \frac{1}{2 S} \partial_\mu \left( S e^\mu_a \right) \gamma^a \Psi+ i m \Psi = 0
\end{equation}
where $S = \left( - \mbox{det} g_{\mu \nu}\right)^{1/2}$ and $(\partial_\mu \Psi)^\sigma = \partial_\mu \Psi^\sigma$.

The above Dirac equation presents several invariances, whatever the space-time dimension may be. First, it is obviously invariant under a change of coordinates since, for each $a$, the numbers $e^\mu_a$ are the components of the vector field $e_a$ and the covariant derivative of a spinor field is, by construction, a co-vector(or $1$-form) field. It is also invariant under an arbitrary, possibly local change of basis in spinor space. Less evidently, any choice of $n$-bein and any choice of operators $\gamma$ obeying the flat space-time Clifford algebra {\sl i.e.} any choice of representation of this algebra, deliver equivalent equations. 

\subsection{II.2 Probability current}

For any space-time dimensions, the probability current conserved by the Dirac dynamics is
\begin{equation}
j^\mu = {\bar \Psi} e^\mu_a \gamma^a \Psi
\end{equation}
where ${\bar \Psi} =  \Psi^\dagger \gamma^0$ is the so-called Dirac conjugate of $\Psi$. 
Let $n$ be the dimension of space-time and consider an arbitrary $n - 1$ dimensional sub-manifold $\Sigma$ with normal $d\Sigma_\mu$. The expression of 
$d\Sigma_\mu$ is
\begin{equation}
d \Sigma_\mu = S\, \frac{1}{(n-1)!} \  \epsilon_{\mu \nu ... \alpha}  dx^\nu \wedge ....\wedge dx^\alpha
\end{equation}
where $\epsilon_{\mu \nu ... \alpha}$ is the completely antisymmetric symbol of degree $n$ with $\epsilon_{0, 1... n} = 1$ and $S$ stands, as above, for the square root of $(- 1)^{n-1}$ times the determinant of the metric components in the coordinate basis. One can form out of $j^\mu$ and $d\Sigma_\mu$ the scalar $dN = j^\mu d\Sigma_\mu$, which represents the probability of finding the Dirac particle in the infinitesimal spatial domain corresponding to $d\Sigma_\mu$. 
Suppose now the sub-manifold $\Sigma$ coincides with a surface of constant $x^0$ with $x^0$ time-like. The only non vanishing component of $d\Sigma_\mu$ is 
$d\Sigma_0 = S dx^1 \wedge ... dx^n = S d^{n-1}x$. It follows that 
\begin{eqnarray}
dN & = & j^0 d\Sigma_0 \nonumber \\
& = & \left( S {\Psi}^\dagger \gamma^0 e^0_a \gamma^a \Psi \right) d^{n-1}x
\end{eqnarray}
so the probability density $P_L$ with respect to the usual Lebesgue measure $d^{n-1}x$ is $P = S {\Psi}^\dagger \gamma^0 e^0_a \gamma^a \Psi$.


\subsection{II.3 Dirac equation formatted for quantum simulation}


In $(1 + 1)$D space-times, $P_L$ reads
\begin{equation}
P_L = S {\Psi}^\dagger \gamma^0 \left( e^0_0 \gamma^0 + e^0_1 \gamma^1\right) \Psi,
\end{equation}
Choosing a basis $(b_-, b_+)$ in spin space and representing the $\gamma$ operators by the matrices
\begin{equation}
\gamma^0 = 
\begin{pmatrix}
0 & 1 \\
1 & 0 
\end{pmatrix}
\end{equation}
and
\begin{equation}
\gamma^1 = 
\begin{pmatrix}
0 & 1 \\
 -1 & 0 
\end{pmatrix},
\end{equation}
the probability density $P_L$ takes the form:
\begin{equation}
P_L = S \left[(e^0_0 - e^0_1) \left(\Psi^-\right)^* \Psi^- + (e^0_0 + e^0_1) \left(\Psi^+\right)^* \Psi^+ \right].
\end{equation}
We have proven above that both $e^0_0$ is strictly superior to both $- e^0_1$ and $e^0_1$. This ensures that $P_L$ is positive definite, as expected from a probability density. 


The above expression for $P_L$ in $(1 + 1)$D space-times suggests introducing the new wave-function components
\begin{eqnarray}
\Phi^- & = &S^{1/2}  (e^0_0 - e^0_1)^{1/2} \Psi^- \\
\Phi^+ & = & S^{1/2} (e^0_0 + e^0_1)^{1/2} \Psi^+,
\end{eqnarray}
so that $P_L = \mid \Phi^- \mid^2 + \mid \Phi^+ \mid^2$.
Rewriting the Dirac equation in terms of $\Phi = \Phi^- b_- + \Phi^+ b_+$ delivers
\begin{equation}
\partial_0 \Phi  + N(x) \partial_1 \Phi + \frac{1}{2} (\partial_1 N) \Phi = M(x) \sigma_2 \Phi
\label{eq:DiracQS}
\end{equation}
where 
\begin{equation}
N(x) = 
\begin{pmatrix}
\frac{\sigma^1(x)}{\sigma^0(x)} & 0 \\
0 & \frac{\delta^1(x)}{\delta^0(x)} 
\end{pmatrix}
\end{equation}
with 
\begin{eqnarray}
\sigma^\mu (x) & = & e^\mu_0 (x) + e^\mu_1 (x) \nonumber \\
\delta^\mu (x) & = & e^\mu_0 (x) - e^\mu_1 (x).
\end{eqnarray}
The effective mass is
\begin{equation}
M(x) = m 
\left(\sigma^0(x) \delta^0(x)\right)^{-1/2}
\end{equation} 
and
\begin{equation}
\sigma_2 = 
\begin{pmatrix}
0 & i \\
-i  & 0
\end{pmatrix}
\end{equation}
is the second Pauli matrix.

The relations (\ref{eq:2beinrel}) can be rewritten in terms of the $\sigma$'s and $\delta$'s, and read
\begin{eqnarray}
\sigma^0 \delta^0  & = & g^{00} \nonumber \\
\frac{1}{2}\, \left( \sigma^0 \delta^1 +  \sigma^1 \delta^0\right)& = & g^{01} \nonumber \\
\sigma^1 \delta^1 & = & g^{11}.
\label{eq:2beinrelbis}
\end{eqnarray}
Expressing also the operator $N$ in terms of the $\sigma$'s and $\delta$'s delivers
\begin{equation}
N(x) = \lambda (x) 
\begin{pmatrix}
1 & 0 \\
0 & 1
\end{pmatrix}
+c(x) 
\begin{pmatrix}
- 1 & 0 \\
0 &  1
\end{pmatrix}
\end{equation}
with $\lambda (x) = g^{01}(x)/g^{00}(x)$ and $c(x) = 1/\left(g^{00}(x) S(x)\right)$, where $S(x)$ is the square root of the opposite determinant of the (covariant) metric components. 
Finally, the effective mass $M(x)$ is given by
\begin{equation}
M(x) = m 
\left(g^{00}(x)\right)^{-1/2}
\label{eq:defM}
\end{equation}

\section{III. Generalized QWs}




Let us present these in $(1 + 1)D$
discrete space-time.  Discrete time is indexed by $j \in \mathbb{N}$ and discrete space indexed by $p \in \mathbb{Z}$. The value of an arbitrary quantity $f$ at point $(j, p)$ is denoted by $f_{j, p}$

\subsection{III.1 Spatial derivative operator} 

In equation (\ref{eq:DiracQS}), keep only the time derivative in the l.h.s, and thus move to the r.h.s. the second and third term from the l.h.s. of (\ref{eq:DiracQS}). Apart form the mass term, each component of $\Phi$ then appears as acted upon by a spatial derivative operator $L$ of the form 
\begin{equation}
L (x) = a(x) \partial_1 + \frac{1}{2} (\partial_1 a) {\mathbf 1}
\label{eq:defa}
\end{equation}
with $a(x) = - \lambda(x) + c(x) = a^-(x)$ for $\Phi^-$ and $a(x) = - \lambda(x) - c(x) = a^+(x)$ for $\Phi^+$. 

Consider therefore an arbitrary, possibly time-dependent operator $L(x)$ of the form (\ref{eq:defa}) acting on single component wave-functions which are $L^2$ in $x^1$. Introduce also a real positive parameter $\epsilon$. A direct computation shows that $L^\dagger(x) = - L(x)$ so the operator $ {\mathbf 1} + \epsilon L(x)$ is unitary
at first order in $\epsilon$. 
Note that this operator is actually the first order expansion of $\exp(i \epsilon L)$.
 Write now $x^0 = j \epsilon$, $x^1 = p \epsilon$, $(j, p) \in \mathbb{N} \times \mathbb{Z}$
and pick an arbitrary discretisation $(L_D)_{j, p}$ of $L$ {\sl i.e.} an operator ${L_D}_{j, p}$ which acts on functions of $p \in \mathbb{Z}$ and which is equivalent to $\epsilon L$ when 
$\epsilon$ tends to zero. 

The operator $(L_D)_{j, p}$ does not necessarily makes $(V_{L_D})_{j, p} = 1 + (L_D)_{j, p}$ unitary. If it does not, introduce $({\bar L}_D)_{j, p} = \left( (L_D)_{j, p} - (L^\dagger_D)_{j, p} \right)/2$ and consider $({\bar V}_{L_D})_{j, p} = \exp\left(({\bar L}_D)_{j, p}\right)$. Since $({\bar L}^\dagger_D)_{j, p} = - ({\bar L}_D)_{j, p}$, the operator $({\bar V}_{L_D})_{j, p}$ is unitary and its expression at first order in $\epsilon$ is 
$1 + (\epsilon L(x) - (- \epsilon L(x)))/2 = 1 + \epsilon L(x)$. 

Let $(U_{L_D})_{j, p}$ be either $(V_{L_D})_{j, p}$ or $({\bar V}_{L_D})_{j, p}$ if both are unitary, or $({\bar V}_{L_D})_{j, p}$ if $(V_{L_D})_{j, p}$ is not. Consider then then unitary discrete quantum automaton defined on $\mathbb{N} \times \mathbb{Z}$ by the evolution equation
\begin{equation}
\Psi_{j + 1, p} = \left( U_{L_D} \Psi_j \right)_{j, p}.
\end{equation}
Suppose that, for all $f$, $f_{j, p}$ is the value taken by a function $f(x^0, x^1)$ at point $x^0_{j, p} = j \epsilon$ and $x^1_{j, p} = p \epsilon$. The left-hand side of ... can then be written 
as $\Psi (x^0_{j, p} + \epsilon, x^1_{j, p})$ and expanding this at first order in $\epsilon$ delivers
$\Psi (x^0_{j, p} + \epsilon, x^1_{j, p}) =  \Psi (x^0_{j, p}, x^1_{j, p})+ \epsilon {\partial_0 \Psi}_{j, p}$. 
By construction, the expansion of $(U_{L_D})_{j, p}$ at the same order is $(U_{L_D})_{j, p}  = \mathbf{1} + \epsilon L(x_{j, p})$. So, at first order in $\epsilon$, the equation of motion of the automaton transcribes into $\partial_0 \Psi = L(x) \Psi$. 

\subsection{III.2 Definition of the walks} 

Let $L^\pm (x)= a^\pm(x) \partial_1 + \frac{1}{2} (\partial_1 a^\pm) {\mathbf 1}$ be the spatial derivative operators associated to both components of $\Phi$ and let $(L^\pm_D)_{j, p}$ be two discretisations of these operators. Let $(U^\pm_{L_D})_{j, p}$ be the corresponding discrete unitaries and define an operator $({\mathcal U})_{j, p}$ acting on the two component wave-function by
\begin{equation}
{\mathcal U} = 
\begin{pmatrix}
(U^-_{L_D})_{j, p} & 0\\
0 & (U^+_{L_D})_{j, p}
\end{pmatrix}.
\end{equation}
Introduce also
\begin{equation}
R(\theta) = 
\begin{pmatrix}
 \cos \theta & - i \sin \theta \\
 i \sin \theta & -  \cos \theta
\end{pmatrix}
\end{equation}
where $\theta$ is an arbitrary real number. Finally, given an arbitrary collection $\bar \theta$ of time- and position-dependent angles $\theta_{j, p}$, call ${\mathcal R} ({\bar \theta})$ the operator
defined by
\begin{equation}
\left({\mathcal R} ({\bar \theta}) \Phi \right)_{j, p} = R(\theta_{j, p}) \Phi_{j, p}.
\end{equation}
The generalized QWs are then defined by the unitary evolution operator ${\mathcal R}({\bar \theta}_M) {\mathcal U}$ where $\left({\bar \theta}_M\right)_{j, p} = \epsilon M(j \epsilon, p \epsilon)$, with $M$ defined by above in equation (\ref{eq:defM}).

\section{IV. Examples}

\subsection{IV.1 Flat space-time in Minkovski coordinates}

This is clearly the simplest situation possible. The choice $e_0 = \partial_0$ and $e_1 = \partial_1$ leads to $a^- (x) = - a^+ (x) = 1$ and $M(x) = m$. A first natural discretisation $L_D^\pm$ of $L^\pm$ is
\begin{eqnarray}
(L_D^- \Phi^-)_{j, p} & =  & \Phi^-_{j, p+1} - \Phi^-_{j, p} \nonumber\\
(L_D^+ \Phi^+)_{j, p} & =  & \Phi^+_{j, p-1} - \Phi^+_{j, p}.
\end{eqnarray}
The operators $\mathbf{1} + L_D^\pm$ are then both unitary, so one can choose them as $U^\pm_{L_D}$ and the resulting walks coincide with standard free massive Dirac quantum walks in flat space-time. 
But choosing $U^\pm_{L_D} = {\bar V}^\pm_{L_D} = \exp({\bar L}^\pm_D) = \exp\left(\left(L^\pm_D - (L^\pm_D)^\dagger \right)/2\right)$ is also possible. The generalised walks are then different from standard Dirac quantum walks. Indeed, in Fourier ($k$-)space, the shift operator of standard DQWs acting on $\Phi^\pm$ amounts to a simple multiplication by $\exp(-\pm i k)$ but the shift operator of the generalised walks amounts to a multiplication by $\exp(-\pm i \sin(k))$, which is different, but delivers the same continuous limit obtained for $k = \epsilon \rightarrow 0$.

%

Another natural discretisation is
\begin{eqnarray}
(L_D^- \Phi^-)_{j, p} & =  & \Phi^-_{j, p+1} - \Phi^-_{j, p} \nonumber\\
(L_D^+ \Phi^+)_{j, p} & =  & - \Phi^+_{j, p+1} + \Phi^+_{j, p}.
\end{eqnarray}
The operator ${\mathbf 1} + L^-_D$ is unitary, so both choices of $U^-_{L_D}$ are possible, but the operator ${\mathbf 1} + L^+_D$ is {\sl not} unitary, so only the second option remains. This leads to a shift operator for $\Phi^+$ which, in Fourier space, amounts to a multiplication by $\exp(- i \sin(k))$. Choosing the first option for $U^-_{L_D}$ then delivers hybrid walks in which the components of the wave-function are treated asymmetrically by the generalised shift operator. These hybrid walks have nevertheless the same continuous limit as conventional DQWs and may be easier to realise in some experimental set-ups.

\subsection{IV.2 Weak gravitational fields}

In the post Newtonian approximation, $(1 + 3)$D relativistic gravitational fields obey the so-called Gravitoelectromagnetism theory \cite{Wald84a,Mashhoon08a} and can be described by a metric of the form
\begin{equation}
ds^2 = (1 - 2 V) d(x^0)^2 + 4({\bf A}.d{\bf r}) dx^0 + (1 + 2V) \eta_{ij} dx^i dx^j 
\end{equation}
where ${\bf r} = (x^1, x^2, x^3)$ and $\eta_{ij} = \mbox{diag} (-1, -1, -1)$. The function $V$ is a gravitational scalar potential and $\bf A$ a gravitational vector potential. Both $V$ and $\bf A$ are supposed to be much smaller than unity and the gauge invariance takes the form
\begin{eqnarray}
V & \rightarrow & V - \frac{\partial F}{\partial t} \nonumber \\
{\bf A} & \rightarrow & {\bf A} +  \nabla F
\end{eqnarray}
with $F$ being an arbitrary function obeying $\Box F = 0$ where $\Box$ is the usual D'Alembert operator. 

Following the analogy with Maxwell electromagnetism, there are two simple gauge choices to describe a constant gravitational field, say in the $x^1$ direction. The first choice is $V(x^0, x^1) = - g x^1$ and ${\bf A} (x^0, x^1) = 0$ and the second one is $V(x^0, x^1) = 0$ and ${\bf A} (x^0, x^1) = - g x^0$. This second gauge is actually more convenient than the first, because the only non-constant metric component is then $g_{01} = - 2 g x^0$, which depends only on time, thus making it possible to use spatial Fourier transform to study motions in this gauge.

We therefore retain this gauge and study the motion of Dirac fermions in the $2D$ space-time manifold with line-element
\begin{equation}
ds^2 = d(x^0)^2 - 4 g x^0 dx^0 dx^1- (dx^1)^2.
\label{eq:metricGEM}
\end{equation}
The metric components are $g_{00} (x) = - g_{11}(x)  = 1$ and $g_{01} (x) = - 2 g x^0$. One finds $S(x) =  \left( 1 + 4 g^2 (x^0)^2 \right)^{1/2} = \cosh \theta(x)$ with $\sinh \theta(x) = 2 g x^0$. The inverse metric components are 
$g^{00} (x) = - g^{11}(x)  = 1/S^2(x) = 1/\cosh^2 \theta(x)$ and $g^{01} (x) = - 2 g x^0/S^2(x) = -\left( \sinh \theta(x)\right)/\cosh^2 \theta(x)$. It follows that $\lambda(x) = - 2 g x^0 = -\sinh \theta(x)$ and $c(x) = - S(x) = \cosh \theta(x)$. The effective mass is $M(x) = m \cosh \theta(x)$. 

The two eigenvalues of the operator $N$ are
\begin{eqnarray}
a^- (x) = \lambda(x) - c(x) & = & - \exp \left(\theta(x) \right) \nonumber \\
a^+(x) =  \lambda(x) +  c(x) & = &  \exp \left(- \theta(x) \right).
\end{eqnarray}
We define therefore two discrete operators $(L_D^-)_j$ and $(L_D^+)_j$ by
\begin{eqnarray}
\left((L_D^-)_j \Phi^- \right)_{j, p} & = &  
\exp\left( \theta_j \right) \left( \Phi^-_{j, p+1} - \Phi^-_{j, p} \right)\nonumber \\
\left((L_D^+)_j \Phi^+ \right)_{j, p} & = &  
- \exp\left( -\theta_j \right) \left( \Phi^+_{j, p} - \Phi^-_{j, p-1} \right)
\end{eqnarray}
where $\sinh \theta_j =  2 g \epsilon j$ with $\epsilon$ is an arbitrary positive real number. The Fourier modes are the eigenvectors of these operators and the $k$-mode is associated to the eigenvalues
$l^\mp_j (k) = \exp(\pm \theta_j) \left( \exp(\pm i k) - 1\right)$. This shows that the operators $\mathbf{1} + (L_D^\pm)_j$ are not unitary, so we choose $(U^\pm_{L_D})_j =  \exp\left(\left((L^\pm_D)_j - (L^\pm_D)_j^\dagger \right)/2\right)$. 
The eigenvectors of $(U^\pm_{L_D})_j$ are also the Fourier modes and the eigenvalues are $u_j^\mp (k) = \exp \left(\pm  i \exp(\pm \theta_j) \sin k\right)$. 

Thus, in Fourier space, the equations defining the generalized walk read:
\begin{eqnarray}
{\hat \Phi}^-_{j + 1} (k) & = & u_j^-(k)  \cos ((\theta_M)_j) {\hat \Phi}^-_j (k) - i u_j^+(k)  \sin ((\theta_M)_j) {\hat \Phi}^+_j (k) \nonumber \\
{\hat \Phi}^+_{j + 1} (k) & = &i  u_j^-(k)  \sin ((\theta_M)_j) {\hat \Phi}^-_j (k) - u_j^+(k) \cos ((\theta_M)_j) {\hat \Phi}^+_j (k)
\end{eqnarray}
where $(\theta_M)_j = \epsilon m \cosh \theta_j$.

Of particular interest is the propagation of massless fermions, which classically follow the null geodesics defined by the equation $ds^2 = 0$. Equation (\ref{eq:metricGEM}) leads to $v^2 + 4 g x^0 v -1 = 0$ where $v = dx^1/dx^0$ is the velocity of a classical massless particle in coordinates $(x^0, x^1)$. Solving this quadratic equation delivers as possible velocities the two celerities $a^\pm$ which enter the Dirac equation, {\sl i.e.} $a^\pm (x^0) = \mp \exp\left(\pm \theta(x^0)\right)$ with $\sinh \theta(x^0) = 2 g x^0$ as before. Integrating one more time gives 
\begin{equation}
(x^1)^\pm (x^0) = \pm \, \frac{1}{4 g}\, \left( \theta(x^0) \pm \frac{\exp\left(\pm 2 \theta(x^0)\right) - 1}{2}\right) + (x^1)^\pm (0).
\label{eq:nullgeod}
\end{equation}
Quantum particles have no trajectory but, by Ehrenfest theorem, the expectation value of the position should obey the non quantum equations of motion and, thus, follow the classical geodesics. 
It can therefore be expected that, for Dirac quantum walks (DQWs), the expectation value of the position operator also follows the classical geodesics, at least for wave-packets which vary  on scales much larger than the discretisation. Figure 1 presents the density plots of the quantum walk for a Gaussian initial condition with variance $\sigma^2 = 300$ much larger than the discretisation $\epsilon = 1$ and with periodic spatial boundary conditions. The gravitational field is $g = -0.2$ and the corresponding null geodesics are plotted as dashed curves.
The figure shows that the density of the quantum does indeed follow the null geodesics, which display a clear bending of the geodesics towards negative values of $x^1$

\begin{figure}[t]
\includegraphics[width=.85\linewidth]{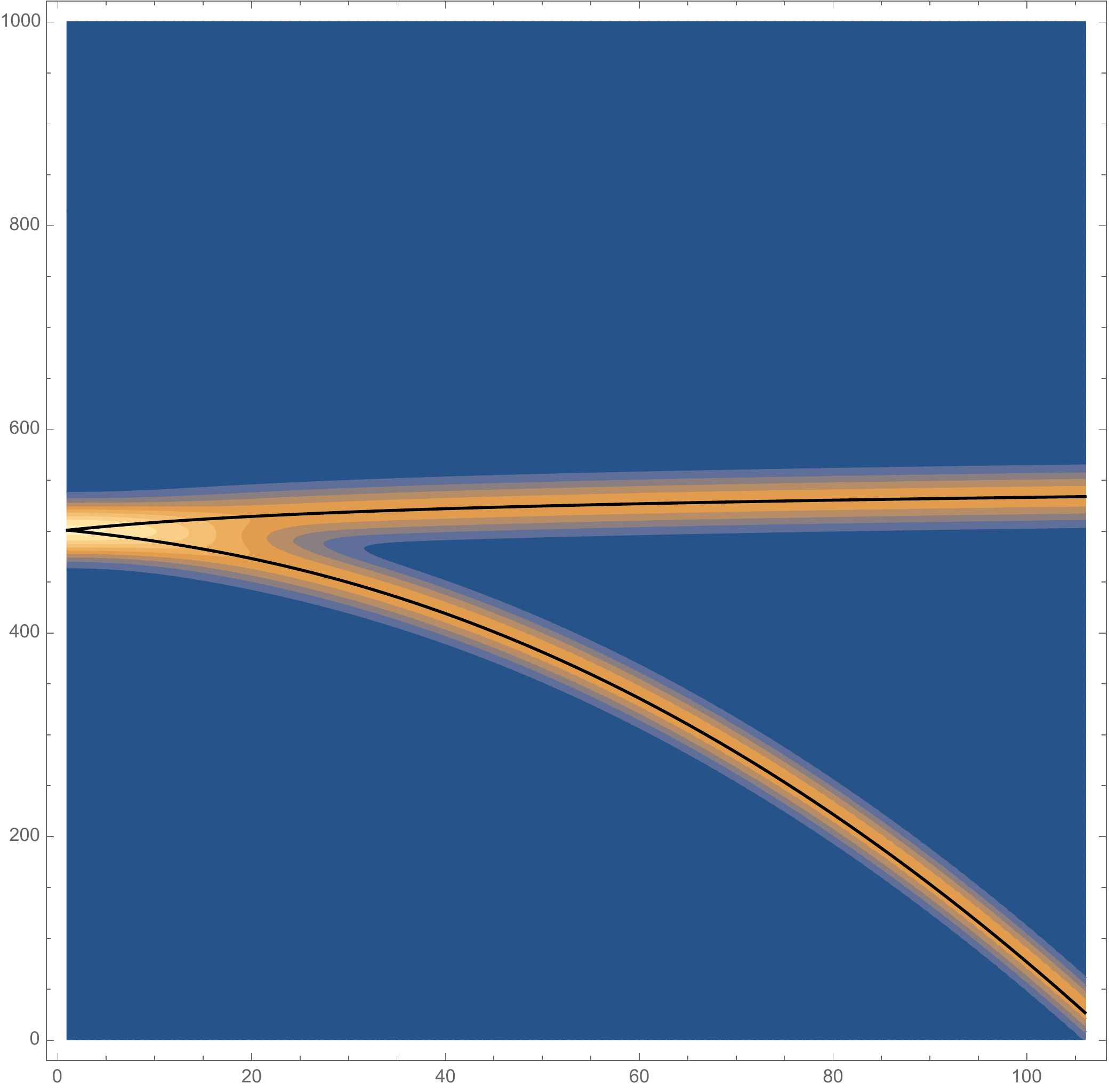}
\includegraphics[width=.1\linewidth]{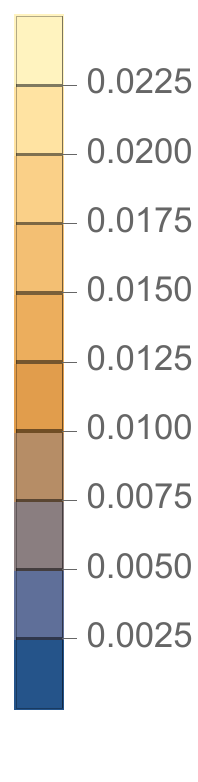}
\caption{Density contour of the quantum walk in discrete time (horizontal) and discrete periodic space (vertical) for vanishing mass. Null geodesics are plotted in black.}
\label{Figure2}
\end{figure}

\section{V. Conclusion}

We have presented new quantum automata, which are built like quantum walks, but with a generalised shift operator, and which admit the $(1 + 1)$D Dirac equation as formal continuous limit. These automata make it possible to simulate solutions the Dirac equation in arbitrary gravitational fields and arbitrary coordinate systems using spinors with two components and not four as previously proposed automata. These new automata are thus minimal and,  therefore, particularly well-suited to NISQ computers. We have also presented as illustration the simulation of massless Dirac fermions in the post-Newtonian approximation.

The above results can and should be extended in various directions. One should first extend the construction of the new automata to simulate Dirac propagation in higher dimensional space-times. Also, the new construction presented in this article start from an arbitrary discretisation on the continuous Dirac equation, and different discretisations will generally result in different automata. Which begs the question: is there a sense in which some of these automata are better than others? And, when standard quantum walks are also an option, can one find initial discretisations which deliver more efficient automata?
Can the new automata be used for spatial search, and/or state transfer, to name just a few applications of standard quantum walks? 

The new automata should also be generalised to include arbitrary Yang-Mills fields. One wonders if the resulting automata will present the same remarkable discrete gauge invariance properties as standard Dirac quantum walks. Can one build discrete field strength `tensors' from the automata? Can they also be used to build a complete discrete geometry, as Dirac quantum walks can? And, if so, what are the differences between the geometries generated by different automata? 

Even more generally, can one build these new quantum automata on graphs? And do they constitute a universal computational primitive, as quantum walks do?

\bibliography{Minimal_Grav_QW_arXiv}
\end{document}